# Low-SWaP magneto-optical trap enabled by planar photonic and magnetic components

Hao Gao[1], Yumeng Zhu[2,3], Zhilong Yu[1,3], Yuhui Hu[1], Zhelin Lin[1], Shiming Wei[2,3], Feng Zhao[2,3], Amit Agrawal[4,5], Zeyang Liu[1*], Xiaochi Liu[2,3*], and Cheng Zhang[1*]

[1]School of Optical and Electronic Information & Wuhan National Laboratory for Optoelectronics, Huazhong University of Science and Technology, Wuhan 430074, China

[2]University of Chinese Academy of Sciences, Beijing 100049, China

[3]Key Laboratory of Atomic Frequency Standards, Innovation Academy for Precision Measurement Science and Technology, Chinese Academy of Sciences, Wuhan 430071, China

[4]Department of Engineering, University of Cambridge, Cambridge CB3 0FA, United Kingdom

[5]Kyung Hee University, 26 Kyungheedae-ro, Dongdaemun-gu, Seoul 02447, Korea

[*]Corresponding author emails: zeyangliu@hust.edu.cn; liuxc@apm.ac.cn; cheng.zhang@hust.edu.cn

**Abstract**

Compact, lightweight, and energy-efficient cold-atom systems are foundational for the development of deployable quantum technologies, yet their realization remains largely constrained due to reliance on bulky optical and magnetic components. Here, we demonstrate a record-low-SWaP (size, weight, and power) magneto-optical trap architecture seamlessly integrating planar photonic and magnetic components into a monolithic, scalable, and manufacturable platform. This is achieved by developing a multifunctional metasurface that converts a linearly-polarized Gaussian beam into a circularly-polarized flat-top beam, replacing conventional lens-waveplate assemblies. In parallel, a planar magnetic coil chip substitutes bulky anti-Helmholtz coils and generates the required quadrupole magnetic field with significantly reduced power consumption. Using $D_2$ line cooling of $^{87}$Rb atoms, the fully planar system achieves nearly an order-of-magnitude improvement in trapped-atom number while operating at a fraction of the SWaP of traditional implementations. This planar integration strategy provides an energy-efficient and scalable pathway toward robust, deployable cold-atom platforms.

## Introduction

Quantum systems based on cold atoms (*1-3*), where an ensemble of atoms are cooled and trapped to near absolute zero temperatures, form the backbone of critical applications including navigation (*4, 5*), time keeping (*6, 7*), quantum computing (*8*), sensing (*9*), and transduction (*10, 11*). Current atom cooling and trapping methods include optical molasses (*12, 13*), dipole traps (*14, 15*), optical lattices (*16, 17*), and magneto-optical traps (MOTs) (*18, 19*). Among them, MOTs combine magnetic field and laser beams to efficiently cool and trap a large population of atoms, and therefore, stand out as one of the most robust and widely used techniques nowadays. In a typical three-dimensional (3D) MOT system, the trap is established within a vacuum environment via scattering forces created by several laser cooling beams and a closed magnetic trap generated by a pair of anti-Helmholtz coils (*20, 21*). Such configuration typically involves three pairs of counter-propagating laser beams of orthogonal circular polarization, alongside a quadrupole magnetic field. To construct such a MOT system, a series of optical components including mirrors, prisms, beam-splitters, and waveplates are used to prepare the laser cooling beams in appropriate polarization states and deliver them along their designed paths. At the same time, anti-Helmholtz coils are commonly employed to generate the quadrupole magnetic field, which inevitably occupy considerable space and add to the weight of the system. Such bulky and intricate configuration presents challenges for integrating MOTs into low-size, weight, and power (SWaP) quantum systems that rely on cold atoms. Consequently, recent efforts have aimed at miniaturization of various MOT components, leading to the development of pyramid-shaped MOTs (*22, 23*), grating MOTs (GMOTs) (*24, 25*), Fresnel MOTs (*26*), and photonic integrated circuit (PIC)-based MOTs (*27, 28*). These elegant approaches allow for the manipulation and delivery of laser beams through miniature or on-chip components, effectively reducing the spatial footprint of laser cooling setups. Among them, the GMOT approach stands out for its robustness and simplicity, where a planar diffraction grating splits a single free-space laser beam into several intersecting beams and thus eliminates the need for generating and aligning multiple discrete laser beams. This technique has demonstrated feasibility across various compact and miniature cold-atom-based quantum instruments such as atomic clocks (*29, 30*), atomic interferometers (*31*), and quantum vacuum standards (*32*).

Although GMOT simplifies laser cooling by eliminating the need for multiple discrete laser beams, it imposes more stringent requirements on the cooling beams compared to a conventional MOT. For example, the schematic of the cooling setup for a four-comb-shaped GMOT system (*33*) using conventional optical and magnetic components is illustrated in Fig. 1A (left panel). The laser beam is conditioned into the required cooling geometry and directed onto the grating chip through a series of conventional optical components. The incident laser beam and the corresponding grating-diffracted beams intersect inside the vacuum cell, achieving equilibrium at the center of the quadrupole magnetic field created by a pair of anti-Helmholtz coils. Due to the periodic structure of the grating chip and the need to balance scattering forces by the incident and diffracted laser beams, the incident laser beam needs to be circular polarized and exhibit a uniform intensity distribution (*34-37*). However, typical output beams from laser sources are linearly polarized and exhibit Gaussian intensity profiles, which would disrupt the scattering force balance and thus degrade the cooling efficiency. To mitigate this issue, additional optical components are used to adjust the beam's polarization state and intensity profile. A quarter waveplate is typically employed to convert the beam's polarization state into circular. At the same time, a lens (or a series of lenses) is used to expand the Gaussian beam so that only its central region—having a more uniform intensity profile—illuminates the grating chip (*33, 38*). While effective, this method inevitably consumes considerable space and results in waste of optical power. Additionally, this approach is also prone to cause imbalance in the scattering forces among the intersecting beams generated by the grating chip (*34*). In parallel, most GMOT systems still rely on conventional anti-Helmholtz coils for the quadrupole magnetic field generation, which also demands substantial space and precise adjustment. Despite recent efforts to reduce the volume and power consumption of conventional anti-Helmholtz coils (*39-42*), most approaches still rely on discrete or three-dimensional structures, resulting in a structural mismatch with miniature cold atom systems. Obviously, these space-consuming optical and magnetic components contradict the GMOT's intended goal of system miniaturization. Addressing this challenge requires the development of alternative optical and magnetic elements that can deliver equivalent or improved performance with higher power efficiency, reduced system footprint, and lower weight.

Here, we present a high-performance, low-SWaP GMOT architecture that eliminates the need for bulky, three-dimensional components with isolated functionalities by seamlessly integrating planar, multi-functional photonic and magnetic elements into a unified platform (Fig. 1A, right panel). Our miniaturization strategy relies on two key innovations: (*i*) a dielectric metasurface that simultaneously performs the functionality of beam shaping and polarization control, thereby replacing separate lens-waveplate assemblies; and (*ii*) a lithographically fabricated planar coil chip that generates the required quadrupole magnetic field, effectively substituting bulky anti-Helmholtz coil pairs. Specifically, we implement a transmission-type silicon (Si) metasurface that transforms a linearly-polarized Gaussian beam ($\lambda$ = 780 nm) into a circularly-polarized flat-top beam (FTB) whose cross-sectional area is precisely matched to the grating chip. The resulting FTB exhibits an intensity uniformity of 78.6%, a root-mean-square intensity variation of 8.0%, and a circular polarization degree of 98.6%. Simultaneously, we develop a 2-mm-thick coil chip consisting of ten stacked layers of coplanar annular coils. The coil chip is fully planar, structurally compatible with the grating chip, and generates a magnetic field gradient of 11.7 Gs/cm at only 0.56 W of power consumption—whereas conventional anti-Helmholtz coils typically require several tens of Watts to achieve similar gradients. Compared to conventional GMOT systems using expanded Gaussian beams, the proposed system nominally reduces the volume (weight) of the core optical and magnetic components by 10× (10×) and 1000× (100×), respectively. At the same time, it achieves 3.5× higher trapped-atom count and captures up to $(8.15 \pm 0.15) \times 10^6$ $^{87}$Rb atoms at 100 mW of incident optical power, demonstrating remarkably improved trapping efficiency. This fully planar integration strategy provides a practical and lithography-friendly pathway towards the development of low-SWaP cold-atom systems, with strong potential for miniature or chip-scale atomic clocks, field-deployable interferometers, and space-based quantum instruments.

## Results

### Moving from conventional GMOT systems to low-SWaP GMOT systems

The proposed low-SWaP GMOT system integrates two key planar components: a dielectric metasurface (Fig. 1B, top panel) and a planar coil chip (Fig. 1B, bottom panel). Before we elaborate on the details of these two planar components, we will analyze and compare the scattering forces exerted on $^{87}$Rb atoms by a Gaussian beam, as typically employed in a conventional GMOT system, and by a flat-top beam, as employed in this study.

The grating chip for single laser beam cooling typically consists of three linear gratings symmetrically arranged at 120° angles to form a triangular configuration (Fig. 1B, middle panel). The grating is periodically etched, where each etched groove has identical structural parameters. The scattering forces exerted by the single incident beam and three diffracted beams are denoted as $F_i$ and $F_{d_j}$ (where $j = 1, 2, 3$), respectively. The angles between $F_{d_j}$ and the $z$-axis is equal to the grating diffraction angle $\theta_a$. The components of $F_{d_1}$, $F_{d_2}$, and $F_{d_3}$ in the horizontal plane (the *x-y* plane) are separated by an angle of 120° with respect to each other. In this comparative study, the spatial intensity distribution of the beam illuminating over the square area of the grating chip (with a side length of 20 mm) is set to be either a flat-top profile (with a side length of $L$ = 20 mm, identical to that of the grating chip) or a Gaussian profile (with a beam diameter of $d$ = 20 mm). Here, $d$ is defined as twice the distance from the central axis of the Gaussian beam to where its intensity drops to $1/e^2$ (≈ 13.5%) of the maximum value. In the subsequent comparison, the total energy of the incident beam irradiated on the grating chip is set to be the same, regardless of the beam profile or size. The detailed scattering force analysis is elaborated in Section I, Supplementary Materials.

Figure 1C shows the spatial beam intensity distribution in the *y-z* plane (perpendicular to the grating chip), resulting from the interaction of incident and diffracted laser beams. When the incident beam has a flat-top profile, the GMOT region exhibits a uniform spatial intensity distribution across the *y-z* plane (Fig. 1C, left panel). In contrast, when the incident beam has a Gaussian profile, the GMOT region exhibits a characteristic intensity decay across the *y-z* plane, with beam intensity diminishing as a function of both radial distance from the grating chip center along the *y*-axis and vertical height above the grating chip surface along the *z*-axis (Fig. 1C, right panel).

We now examine the scattering force distribution under the two illumination configurations. In the $x$-$y$ plane, the forces along the positive and negative halves of the $x$- and $y$-axes remain equal in magnitude and opposite in direction about the origin, independent of the illuminating beam profile or size (see Section I, Supplementary Materials). In contrast, the scattering force distribution in the $y$-$z$ plane displays a pronounced difference between the two configurations (Fig. 1D). When the incident circularly-polarized beam has a flat-top profile, the force distribution is nearly centrosymmetric with respect to the coordinate origin and exhibits a zero value at the center in the $y$-$z$ plane (Fig. 1D, left panel). In contrast, when the incident beam has a Gaussian profile, the force is radially asymmetric with respect to the coordinate origin (Fig. 1D, right panel). A pronounced force imbalance along the positive and negative halves of the $z$-axis results in a significant downward shift of the atomic cloud centroid. Simultaneously, the radial force along the $y$-axis becomes negligible near the center of the GMOT, leading to a substantial decrease in the number of trapped atoms. By expanding the Gaussian beam, the force distribution gradually approaches centrosymmetry about the coordinate origin. Notably, when the beam is expanded to $d$ = 60 mm, the force distribution becomes comparable to that of the flat-top beam, thereby providing a favorable condition for atom trapping (details elaborated in Section I, Supplementary Materials).

The above analysis shows that GMOT systems using Gaussian beams require substantial beam expansion to achieve a nearly uniform intensity profile over the grating chip for efficient atom trapping. However, expanding the Gaussian beam demands significant space along its propagation path and leads to substantial laser energy waste. For instance, in the case of a Gaussian beam with a diameter of $d$ = 60 mm illuminating a grating chip with a side length of $L$ = 20 mm, only approx. 20% of the beam energy is utilized for trapping. In contrast, flat-top beams overcome the limitations of Gaussian beams by enabling near-perfect laser energy utilization. Furthermore, for the same laser energy applied to the grating chip, flat-top beams provide a more balanced scattering force distribution and improve atom trapping efficiency. Existing methods for generating flat-top beams, such as using axicons (*43*) and freeform lenses (*44*), are effective but require considerable space and lack flexibility in customizing the beam size and position—both critical for tailored low-SWaP atom trapping systems. In this work, we demonstrate that a single planar optical metasurface can be engineered to generate a flat-top beam over a short propagation distance, with its size precisely tailored to match the grating chip. This approach successfully resolves the space and customization limitations of conventional optical elements. Additionally, the metasurface is dual-functional: it not only shapes the Gaussian beam into a flat-top profile but also simultaneously converts its linear polarization state into circular, further eliminating the need for quarter waveplates used in conventional setups.

In a GMOT cooling architecture, the atoms are cooled and trapped by the combination of optical and magnetic fields. Traditionally, a pair of anti-Helmholtz coils is used to form a magnetic potential well to trap the atoms, which typically consists of two identical coils (with the same radius, current, and number of turns) that carry currents in opposite directions (operational principle of the anti-Helmholtz coils is elaborated in Section II, Supplementary Materials). To achieve a magnetic field gradient exceeding 10 Gs/cm (a typical requirement in alkali-atom trapping systems) (*45*), the required coil currents grow substantially with increasing trap dimensions, creating significant heat dissipation challenges in scaled-up configurations. In this work, we propose a planar coil chip architecture that offers significantly reduced weight and volume by employing a single coplanar annular structure (Fig. 1B, bottom panel). This design incorporates multiple nested coils with varying radii, turn numbers, and current directions, optimized to achieve the desired magnetic field with minimal power consumption. This planar structure also facilitates integration with other on-chip laser cooling components, enhancing compatibility and miniaturization of the whole GMOT system.

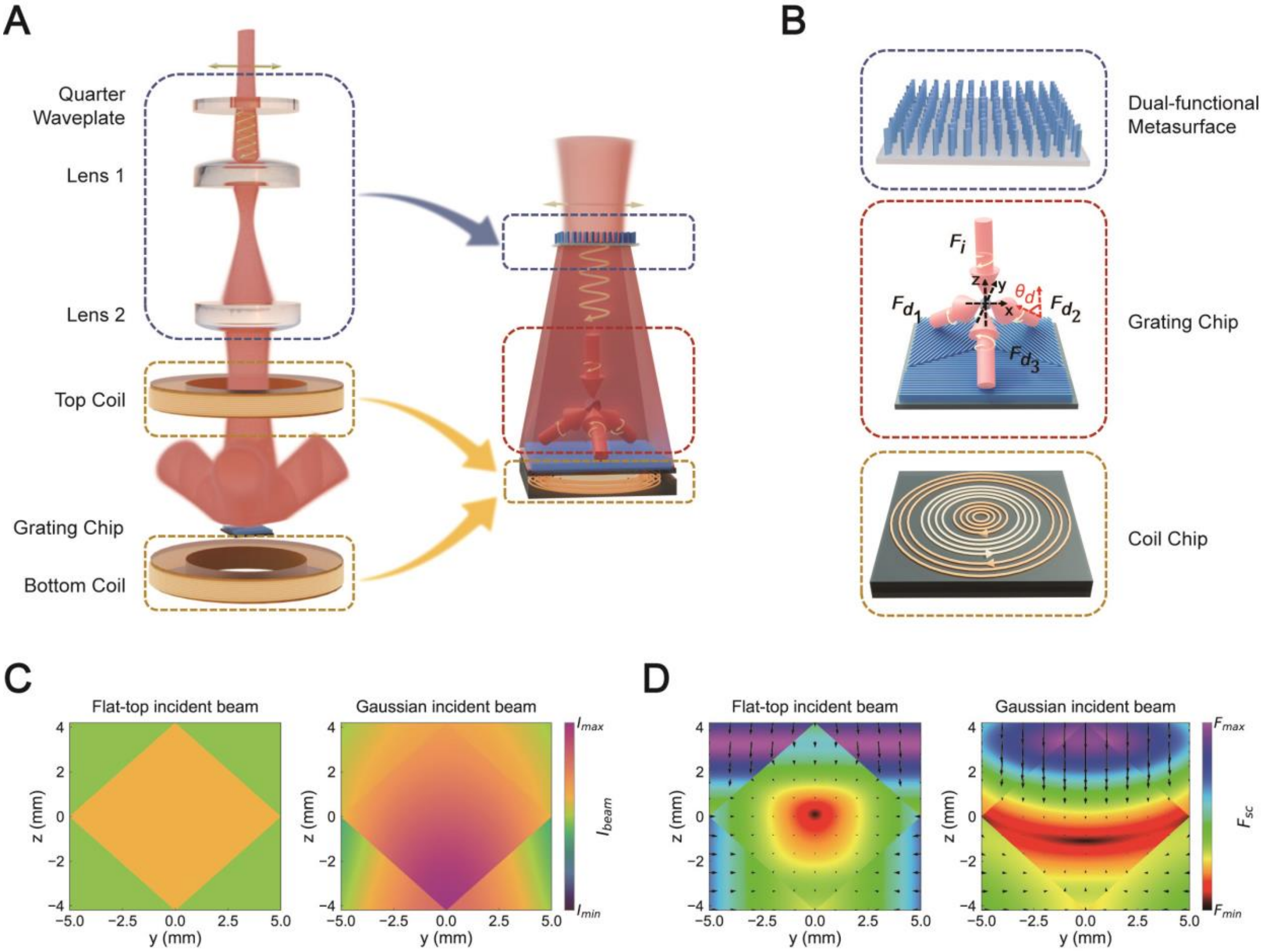

**Fig. 1. Transition from conventional GMOT systems to low-SWaP GMOT systems incorporating both planar photonic and magnetic components. (A)** Left panel: Schematic of a conventional GMOT system utilizing bulky optical and magnetic components. A linearly-polarized Gaussian beam is converted to circular polarization by a quarter waveplate, and gets expanded by a lens pair before reaching the grating chip. The incident laser beam and the three grating-diffracted beams intersect within the vacuum cell, achieving equilibrium at the center of the quadrupole magnetic field generated by a pair of anti-Helmholtz coils. Right panel: Schematic of the proposed low-SWaP GMOT system utilizing a planar dielectric metasurface and a planar coil chip. The metasurface converts the incident linearly-polarized Gaussian beam into a circularly-polarized flat-top beam of side length precisely

matching the grating chip at a short propagation distance beyond the metasurface. The planar coil chip replaces the conventional 3D anti-Helmholtz coils and generates the required quadrupole magnetic field. **(B)** Schematics of the key components in the low-SWaP GMOT system. Top panel: Dual-functional metasurface consisting of high-aspect-ratio Si nanopillars of rectangular cross-section. Middle panel: Si grating chip comprising three linear gratings symmetrically arranged at 120° with respect to each other. Bottom panel: Planar coil chip consisting of ten layers of nested coils in a serial configuration. **(C)** Calculated beam intensity distribution in the *y-z* plane when the grating chip is illuminated by incident beams with flat-top and Gaussian profiles at the same laser energy. Left panel: Illumination with a flat-top profile beam of $L$ = 20 mm, matching the grating chip size. Here, the trapping region exhibits a uniform spatial intensity distribution. Right panel: Illumination with a Gaussian profile beam of $d$ = 20 mm. Here, the trapping region exhibits a gradual intensity decay with increasing radial distance from the grating-chip center and with increasing vertical height above the chip surface. **(D)** Calculated scattering force shown as vector and magnitude in the *y-z* plane for two illumination conditions. Left panel: Under flat-top beam illumination, the force is nearly centrosymmetric about the coordinate origin and exhibits a zero value at the center. Right panel: Under Gaussian beam illumination, the force is radially asymmetric about the coordinate origin, exhibiting an imbalance along the positive and negative halves of the *z*-axis while becoming negligible along the y-axis near the center.

## Dual-functional metasurface for simultaneous flat-top beam generation and linear-to-circular polarization conversion

We design a transmission-type dual-functional metasurface that transforms a linearly-polarized Gaussian beam into a circularly-polarized flat-top beam (FTB), with the output beam size precisely matched to that of the grating chip (Fig. 2A, left panel). Optical metasurfaces, composed of subwavelength nanostructures, offer a powerful platform for multi-dimensional control of light using a planar architecture (*46-62*). In this study, the metasurface is implemented to simultaneously shape the wavefront and control the polarization state of the incident cooling beam, eliminating the need for separate lenses and waveplates typically required in conventional GMOT systems and thereby offering a compact and efficient beam manipulation solution. The electromagnetic response of a metasurface's constituent meta-atom can be represented by the Jones matrix:

$$J = R(-\theta)J_0(o,e)R(\theta) = \begin{bmatrix} \cos\theta & -\sin\theta \\ \sin\theta & \cos\theta \end{bmatrix} \begin{bmatrix} t_o e^{i\Delta_0} & 0 \\ 0 & t_e e^{i\Delta_e} \end{bmatrix} \begin{bmatrix} \cos\theta & \sin\theta \\ -\sin\theta & \cos\theta \end{bmatrix} \quad (1)$$

Here, $\Delta_o$ and $\Delta_e$ are the phase shifts imparted on electric fields along the two main axes (o- and e-axis) of the meta-atom, and $t_o$ and $t_e$ represent the amplitude transmission coefficients of the electric fields along these two axes. $\theta$ denotes the in-plane rotation angle of the meta-atom within its unit cell. To achieve efficient linear-to-circular polarization conversion, the meta-atom needs to function as a quarter waveplate. This can be achieved by setting the difference between the two phase shifts $\Delta_o$ and $\Delta_e$ to be $\pi/2$ ($\Delta = \Delta_o - \Delta_e \approx \pi/2$), and the amplitude transmission coefficients $t_o$ and $t_e$ to be nearly unity ($t_o \approx 1,\ t_e \approx 1$). In the case when the incident light is linearly polarized along the *x*-axis ($E_{in} = [1,0]^T$) and the output light is designed to be left-handed circularly polarized (LCP), the polarization direction of the incident light should make an angle of 45° with respect to the long axis of the meta-atom ($\theta = 45°$). The output electrical field can be represented as:

$$\begin{aligned} E_{out} &= JE_{in} \\ &= e^{i\Delta_o} R(-45°) \begin{bmatrix} 1 & 0 \\ 0 & -i \end{bmatrix} R(45°) \begin{bmatrix} 1 \\ 0 \end{bmatrix} \\ &= \frac{(1+i)e^{i\Delta_o}}{2} \begin{bmatrix} 1 \\ i \end{bmatrix} \end{aligned} \quad (2)$$

It can be observed that the phase of the output LCP light is closely related to the phase shift $\Delta_o$ provided by a given meta-atom. By adjusting the structural parameters of the meta-atoms to provide the required $\Delta_o$ values at different spatial positions across the metasurface plane, an on-demand phase modulation profile can be readily implemented.

In this study, a Gaussian beam is incident on a square-shaped metasurface and subsequently gets transformed into an FTB at a pre-set propagation distance. Based on the energy mapping principle, we derive the phase modulation profile $\phi(x_0, y_0)$ of the metasurface (details elaborated in Section III, Supplementary Materials).

$$\phi(x_0, y_0) = \phi_x(x_0) + \phi_y(y_0) \tag{3}$$

$$\phi_x(x_0) = \int_0^{x_0} \frac{2\pi}{\lambda} \frac{1}{\sqrt{1 + \frac{{z_0}^2}{\left[\frac{\frac{L_x}{2}\operatorname{erf}\left(\frac{\sqrt{2}x}{w_i}\right)}{\operatorname{erf}\left(\frac{\sqrt{2}D_0}{2w_i}\right)} - x\right]^2}}} dx \tag{4}$$

$$\phi_y(y_0) = \int_0^{y_0} \frac{2\pi}{\lambda} \frac{1}{\sqrt{1 + \frac{{z_0}^2}{\left[\frac{\frac{L_y}{2}\operatorname{erf}\left(\frac{\sqrt{2}y}{w_i}\right)}{\operatorname{erf}\left(\frac{\sqrt{2}D_0}{2w_i}\right)} - y\right]^2}}} dy \tag{5}$$

Here, $z_0$ is the propagation distance from the metasurface exit-surface, where the target FTB is formed. $\lambda$ is the free-space wavelength of the incident light, $w_i$ is the waist of the incident Gaussian beam, and $D_0$ is the side length of the square-shaped metasurface. $L_x$ and $L_y$ denote the side lengths of the rectangular-shaped FTB along the *x*- and *y*-axes, respectively. In general, the generated FTB can have different sizes along the *x*- and *y*-axes. In this work, the FTB is designed to match the square-shaped grating chip ($L_x = L_y = L$). The term "erf" denotes the error function, defined as $\operatorname{erf}(q) = \frac{2}{\sqrt{\pi}} \int_0^q \exp(-t^2)\, dt$.

For the constructed GMOT system, the Gaussian beam from the laser source has a free-space wavelength of $\lambda$ = 780 nm (for cooling $^{87}$Rb atoms) and a waist of $w_i = 250$ μm. The size of the generated FTB is designed to match the dimension of the grating chip with a side length of $L = 20$ mm. The propagation distance for FTB formation beyond the metasurface is set as $z_0 = 10$ cm, which ensures a compact system footprint. To fully utilize the incident laser energy, the metasurface is designed to have a side length of $D_0 = 1.5$ mm, three times larger than the incident Gaussian beam diameter. The influence of the metasurface size on the FTB shaping performance is discussed in Section IV, Supplementary Materials. Using these parameters, the phase modulation profile $\phi(x_o, y_o)$ is derived and plotted in Fig. 2B. For the final metasurface design, an additional gradient phase $\phi_g = \frac{2\pi}{\lambda} x_0 \sin\alpha$ is introduced, causing an $\alpha = 8°$ tilt of the FTB along the $x_0$-axis. This phase modification effectively separates the generated FTB from the directly transmitted residual beam and ensures a high beam purity.

To implement the designed dual-functional metasurface, amorphous silicon (a-Si), which exhibits high refractive index and low absorption coefficient at the system's operational wavelength, is chosen as the metasurface's constituent material. The experimentally measured refractive index and absorption coefficient of the a-Si film are displayed in Section V, Supplementary Materials. Each a-Si meta-atom acts as a miniature quarter waveplate for linear-to-circular polarization conversion, and at the same time, the phase of the transmitted circularly-polarized light is controlled by the meta-atom dimensions. To determine the structural parameters of the a-Si meta-atoms, the amplitude transmittance and phase shift for propagation of $\lambda$ = 780 nm light, linearly polarized either (*i*) parallel to the major axis ($t_o$ and $\Delta_o$), or (*ii*) parallel to the minor axis ($t_e$ and $\Delta_e$) of the rectangular a-Si nanopillar are computed using Finite-Difference Time-Domain (FDTD) simulations with periodic boundary conditions. For a chosen pillar height $H$ = 1000 nm and lattice spacing $P$ = 350 nm, the major and minor axis lengths of the pillar, $L_o$ and $L_e$, are iteratively varied to identify orthogonal principal axis combinations simultaneously leading to $\Delta_o - \Delta_e \approx \pi/2$ and $t_o \approx t_e$, in other words, quarter-waveplate-like operation. To facilitate the above parameter search process, a figure-of-merit (FoM) function is defined as $\mathrm{FoM} = \log_{10}\left(\left|\frac{t_o}{t_e} e^{(i(\Delta_o - \Delta_e))} - e^{i\pi/2}\right|\right)$ and displayed in Fig. 2C. The blue-colored regions (i.e., low-FoM-value regions) correspond to various combinations of $L_o$ and $L_e$ that satisfy the target quarter-waveplate-like operation. Details of the nanopillar parameter search process are elaborated in Section VI, Supplementary Materials. Through this process, 1212 a-Si nanopillars with different orthogonal principal axis combinations are selected, providing a full 0 to 2π phase shift coverage with phase intervals down to 0.027 radians. These nanopillars also exhibit high linear-to-circular polarization conversion efficiency $\eta$ exceeding 80%, and a low root-mean-square (RMS) efficiency variation of 4.2% (Fig. 2D). Finally, the dual-functional metasurface is constructed by mapping the target phase modulation profile to the obtained meta-atom library.

The metasurface fabrication process comprises several key steps, including a-Si film deposition via plasma-enhanced chemical vapor deposition (PECVD) on fused silica substrate, electron beam lithography (EBL), chromium (Cr) etching mask lift-off, and inductively coupled plasma reactive ion etching (ICP-RIE). A detailed description of the fabrication process is provided in the Materials and Methods section. The fabricated square-shaped metasurface has a side length of 1.5 mm and visually exhibits uniform color with minimal defects. Scanning electron microscope images show high-aspect-ratio (up to 14:1) a-Si nanopillars with well-defined rectangular cross-sections and smooth sidewalls (Fig. 2A, right panel).

To characterize the device performance, the metasurface is positioned at the beam waist of the Gaussian beam. The intensity distribution of the beam illuminating over the metasurface is displayed in Fig. S8, Supplementary Materials. The intensity profile and polarization state of the transmitted FTB are recorded using a complementary metal-oxide semiconductor (CMOS) camera and a near-infrared polarimeter, respectively. Figure 2E shows the intensity profiles of the generated FTB at different propagation distances beyond the metasurface ($z_0$ = 6, 8, 10, and 12 cm). As $z_0$ increases, the size of the FTB gradually expands, and the beam edges become sharper. Figure 2F displays the cross-sectional cut of the intensity profile at the designed propagation distance of $z_0$ = 10 cm, where the grating chip will be positioned. To assess the quality of the generated FTB, two parameters are defined: the flatness factor $F = \frac{I_{average}}{I_{maximum}}$ and the root-mean-squared (RMS) intensity variation $\delta = \sqrt{\frac{\sum_{i=1}^{n}(I_i - I_{average})^2}{n}}$ (*63*). Here, $I_{average}$ and $I_{maximum}$ are the average and maximum intensity values in the 2D cross-section of the FTB, respectively. $I_i$ is the intensity of the sampling point $i$ and $n$ = 4.94 × $10^6$ is the total number of sampling points in the beam's 2D cross-section. At the designed propagation distance, the square-shaped FTB exhibits a flatness factor $F$ = 78.58% and an RMS intensity variation $\delta$ = 0.08. The device not only achieves comparable RMS intensity variation to existing solutions (*64-66*), but also enables an almost perfect linear-to-circular polarization conversion. The measured polarization states of the incident linearly-polarized Gaussian beam and the transmitted circularly-polarized FTB are mapped on the Poincaré sphere (Fig. 2G). The degree of circular polarization (DOCP), which evaluates the metasurface's polarization conversion capability, is measured to be 98.64%. The metasurface's operational efficiency, defined as the ratio between the intensity of the FTB at the target propagation distance and the intensity of the incident Gaussian beam, is measured to be 42.01%.

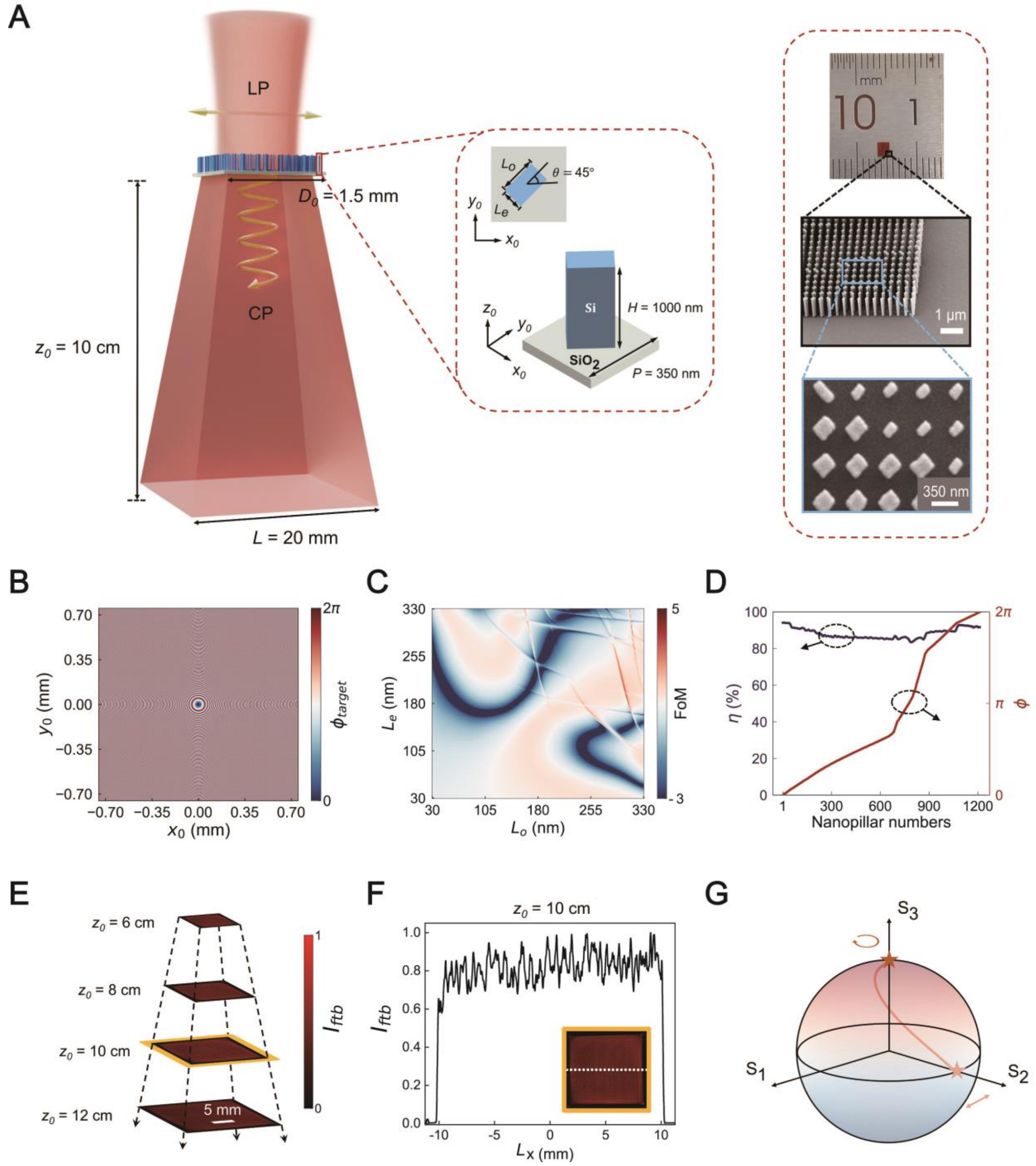

**Fig. 2. Dual-functional metasurface for simultaneous flat-top beam generation and linear-to-circular polarization conversion.** **(A)** Left panel: Schematic of the dual-functional metasurface that transforms an incident linearly-polarized Gaussian beam into a circularly-polarized flat-top beam of size matching the grating chip. Inset: Perspective and top views of the metasurface unit-cell, consisting of an a-Si nanopillar with a rectangular cross-section (side lengths of $L_o$ and $L_e$) and rotation angle $\theta$ = 45° arranged on a $SiO_2$ substrate in a square lattice of pitch $P$. Right panel: Photo, oblique-view (viewing angle: 49°) and top-view SEM images of the fabricated metasurface. **(B)** Phase modulation

profile of the metasurface. **(C)** Quarter-waveplate figure of merit (FoM) versus nanopillar in-plane dimensions ($L_o$ and $L_e$) at the target free-space wavelength $\lambda$ = 780 nm. **(D)** Linear-to-circular polarization conversion efficiency $\eta$ and phase shift $\phi$ of the selected 1212 Si nanopillars. **(E)** Normalized intensity profiles of the generated FTB at propagation distances of $z_0$ = 6, 8, 10, and 12 cm beyond the metasurface. **(F)** Cross-sectional cut of the intensity profile of the FTB at the designated working distance ($z_0$ = 10 cm), where the grating chip is positioned. Inset: Measured two-dimensional intensity profile of the FTB. The dashed white line indicates the sampling direction for the cross-sectional cut. **(G)** Polarization states of the incident linearly-polarized Gaussian beam and the transmitted circularly-polarized FTB mapped onto the Poincaré sphere (denoted by stars).

### Miniaturized planar coil chip for magnetic field generation

The detailed schematic of the planar magnetic coil chip is shown in Fig. 3A. This planar design consolidates the functionality of a conventional anti-Helmholtz coil pair into a single coplanar annular structure, generating a null magnetic field with sufficient gradient that spatially coincides with the overlapping region of the incident and grating-diffracted laser beams. Owing to its simple architecture and lithography-compatible fabrication, the coil chip serves as a direct replacement for conventional anti-Helmholtz coils, enabling planarization of all core components in the GMOT system. Structurally, it comprises multiple nested coils with varying radii, turn numbers, and current directions, patterned on a printed circuit board (PCB), connected in series, and driven by a common current. By appropriately optimizing these parameters, the quadrupole magnetic field can be precisely tailored, with the trapping position flexibly tunable to a desired height above the chip to support compact system integration. To further enhance magnetic field generation efficiency while minimizing power consumption and heat accumulation, multiple layers of coplanar annular coils are vertically stacked and connected in series.

For the system to function properly, the quadrupole magnetic field needs to exhibit a zero value at the laser beam intersection point, and at the same time, maintain a spatial gradient of approximately 12 Gs/cm (*45*). The strength and spatial gradient of the magnetic field along the $z_{coil}$-axis generated by the chip are calculated as:

$$B_{z_{coil}}(0,0,z_{coil}) = \frac{\mu_0}{2}\left( \pm \frac{N_1 {R_1}^2 I}{\left({R_1}^2 + z_{coil}^2\right)^{\frac{3}{2}}} \pm \frac{N_2 {R_2}^2 I}{\left({R_2}^2 + z_{coil}^2\right)^{\frac{3}{2}}} \cdots \pm \frac{N_n {R_n}^2 I}{\left({R_n}^2 + z_{coil}^2\right)^{\frac{3}{2}}} \right) \quad (6)$$

$$\frac{\partial B_{z_{coil}}}{\partial z_{coil}}(0,0,z_{coil}) = \frac{3\mu_0 z_0}{2}\left( \pm \frac{N_1 {R_1}^2 I}{\left({R_1}^2 + z_{coil}^2\right)^{\frac{5}{2}}} \pm \frac{N_2 {R_2}^2 I}{\left({R_2}^2 + z_{coil}^2\right)^{\frac{5}{2}}} \cdots \pm \frac{N_n {R_n}^2 I}{\left({R_n}^2 + z_{coil}^2\right)^{\frac{5}{2}}} \right) \quad (7)$$

Here, $\mu_0$ is the magnetic permeability in vacuum. $I$ is the current applied to the series-connected planar coils. $N_i$ and $R_i$ represent the number of turns and radius of the *i*-th nested coil, respectively. The "+" symbol denotes counter-clockwise coil current flow, while "-" indicates clockwise flow. $n$ is the total number of nested coils. By carefully designing these parameters, we can achieve the target magnetic field with a strength $B_{z_{coil}}$ = 0 and spatial gradient $\frac{\partial B_{z_{coil}}}{\partial z_{coil}} \approx$12 Gs/cm at the target operating height of $H_z$ = 6.5 mm above the chip center. The detailed design process is elaborated in Section VII, Supplementary Materials.

We fabricate the coil chip, comprising ten stacked layers of coplanar annular coil structures, by patterning copper wires on a PCB. Each layer is 70 μm thick, with a wire width of 545 μm and a spacing of 150 μm between adjacent wires. Each layer consists of three nested coil groups, separated from each other by an insulating layer. The innermost coil has $N_1$ = 1 turn with a radius of 1.4 mm. The middle coil comprises $N_2$ = 13 turns with radii ranging from 2.095 mm to 10.435 mm. The outermost coil includes $N_3$ = 11 turns with radii spanning from 11.13 mm to 18.08 mm. Figure 3B shows the photograph of a conventional anti-Helmholtz coil pair (left) and the developed planar coil chip (right), both capable of generating comparable magnetic field gradients. In contrast to the bulky conventional coil, the planar chip has a total thickness of 2 mm and a footprint of 39 × 39 mm—representing a reduction in volume by orders of magnitude. In addition, the weight of the chip is only 8.95 g, more than 100 times lighter than the conventional anti-Helmholtz coil. Figure 3C displays the simulated magnetic field distribution of the coil chip, showing the magnetic field pattern in the $y_{coil}$-$z_{coil}$ plane (perpendicular to the coil plane) under a driving current of $I$ = 2.2 A. At a height of $H_z$ = 6.5 mm, the magnetic field strength approaches 0 Gs, with a spatial gradient reaching 11.86 Gs/cm.

The chip performance is experimentally evaluated. The strength of the magnetic field at different heights above the coil center, under an applied current of 2.2 A, is measured by a magnetometer with a step size of 0.5 mm (Fig. 3D). The chip generates a zero-strength magnetic field at a height of 6.5 mm, with a field gradient of 11.7 Gs/cm along the $z_{coil}$-axis. Figure 3E presents the experimental result of power consumption versus magnetic field gradient. As the coil power increases, the spatial gradient of the magnetic field at the operating point (where the magnetic field strength equals zero) gradually becomes larger. The power consumption, when the field gradient is 11.7 Gs/cm at the operating point, is calculated to be 0.56 W. As a comparison, conventional anti-Helmholtz coils typically consume several tens of Watts of power for realizing similar field gradients (details elaborated in Section II, Supplementary Materials). These results demonstrate that the planar coil chip meets the steady-state magnetic field gradient and power consumption requirements for energy-efficient MOT systems.

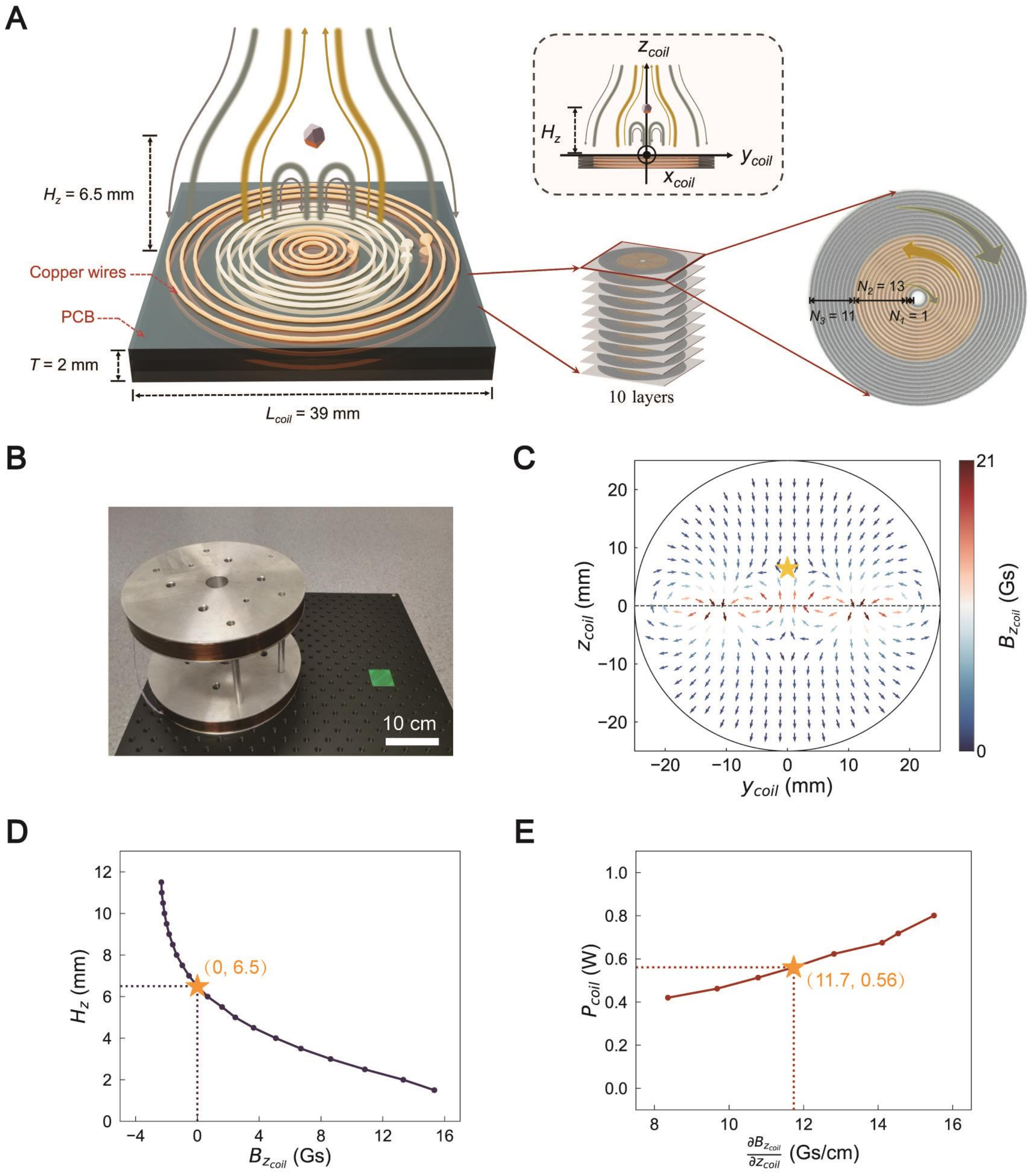

**Fig. 3. Miniaturized planar coil chip for magnetic field generation. (A)** Schematic of the planar coil chip designed to generate the quadrupole magnetic field, which incorporates ten layers of vertically-stacked coplanar annular coil structures. Every layer comprises three groups of nested coils: the innermost coil ($N_1$ = 1 turn, radius = 1.4 mm), the middle coil ($N_2$ = 13 turns, radii ranging from 2.095 mm to 10.435 mm), and the outermost coil ($N_3$ = 11 turns, radii ranging from 11.13 mm to 18.08 mm). The implemented coil chip has a total thickness of 2 mm and a footprint of 39 × 39 mm. Inset: Side view of the planar coil chip. **(B)** Photograph comparing a conventional anti-Helmholtz coil pair (left,

1174.94 g) and the developed planar coil chip (right, 8.95 g). **(C)** Simulated magnetic field distribution in the $y_{coil}$-$z_{coil}$ plane at an applied current of 2.2 A. The arrow direction indicates the orientation of the magnetic field, while the color scale represents the field strength. At the operating point (marked by the yellow star), the magnetic field strength approaches 0 Gs, with a gradient of 11.86 Gs/cm. **(D)** Measured magnetic field strength at different heights from the coil center, under an applied current of 2.2 A. Measurements taken along the $z_{coil}$-direction reveal that the field strength becomes zero at a height of 6.5 mm (indicated by the orange star), with a spatial gradient of 11.7 Gs/cm. **(E)** Measured power consumption versus magnetic field gradient at the zero-field point. As the coil power increases, the magnetic field gradient correspondingly gets larger. For generating a field gradient of 11.7 Gs/cm, the power consumption is 0.56 W (indicated by the orange star).

**Low-SWaP GMOT system incorporating both planar photonic and magnetic components**

The constructed GMOT setup for laser cooling and trapping of $^{87}$Rb atoms is illustrated in Fig. 4A. A 780 nm cooling laser is frequency-stabilized to the $5S_{1/2}$ (F = 2) → $5P_{3/2}$ (F' = 3) cycling transition on the $D_2$ line of $^{87}$Rb atoms using the modulation transfer spectrum (MTS). The frequency is red-detuned by 8.4 MHz to maximize the number of cooled atoms. A 795 nm repumping laser is locked onto the $5S_{1/2}$ (F = 1) → $5P_{3/2}$ (F' = 2) transition of $^{87}$Rb atoms by the saturated absorption spectrum (SAS). Both linearly-polarized laser beams are coupled into a polarization-maintaining fiber and subsequently out-coupled into free space by a fiber collimator. The dual-functional metasurface reshapes the linearly-polarized output Gaussian beam into a circularly-polarized square-shaped FTB which subsequently illuminates the planar grating chip. The FTB, together with the three grating-diffracted beams, form an overlapping region in the vacuum cell. The planar coil chip is positioned beneath the grating, with its center aligned vertically with the overlapping region. Ultimately, the $^{87}$Rb atoms are cooled and trapped within the beam overlapping region, located 6.5 mm above the coil chip.

We now investigate the unique advantages of illuminating the grating chip with a beam having a flat-top intensity profile. As demonstrated by the theoretical analysis in the earlier text, such illumination configuration provides more balanced optical forces on the atoms. In the experiment, the trapping performance of two illumination beams with different profiles is compared: a flat-top profile with a side length $L$ = 20 mm matching that of the grating chip, and a Gaussian profile with varying cross-sectional diameters of $d$ = 40 mm, 50 mm, and 60 mm. In the case of Gaussian beam illumination, a convex lens pair is used to expand the Gaussian beam from the fiber collimator to diameters of 40 mm, 50 mm, and 60 mm (schematic shown in Fig. S11, Supplementary Materials). The effect of atom trapping can be observed from the experimentally measured fluorescence image of the trapped atom cluster (*67*) (details elaborated in Section VIII, Supplementary Materials). When the intensity profile of the beam illuminating the grating chip is set to be flat-top, the trapped-atom cluster exhibits a circular shape (Fig. 4B, top panel), in contrast to the flattened shape observed when the beam has a Gaussian profile (Fig. 4B, bottom panel). As the laser power increases, significant differences are observed in the growth trend of trapped-atom number between the two cases. For the case of FTB illumination, the number of trapped atoms increases linearly with laser power, reaching a maximum of $(8.15 \pm 0.15) \times 10^6$ atoms at an incident power of 100 mW (Fig. 4C, red curve). The cited uncertainty represents one standard deviation of the measured data. This follows the typical trend observed in atom number variation under low atomic transitions (*68*). It is worth noting that the aforementioned laser power corresponds to the maximum output from the employed laser system. With even higher laser powers, the trapped-atom number is expected to continue to rise before eventually plateauing due to the saturation of atomic transitions. In sharp contrast, for the case of Gaussian beam illumination, the number of trapped atoms only increases slightly with laser power before plateauing. As the diameter of the expanded Gaussian beam increases, more atoms can be trapped at the same power level due to the more uniform intensity distribution over the grating area and thus better atom trapping efficiency. But still, the trapped atom numbers for all three Gaussian beams are consistently lower than that of the FTB under the same illumination power. As an example, the trapped-atom numbers at the laser power of 100 mW are measured to be $(2.30 \pm 0.31) \times 10^6$, $(3.91 \pm 0.12) \times 10^6$, and $(4.23 \pm 0.13) \times 10^6$ respectively for the Gaussian beams having diameters of 40 mm, 50 mm, and 60 mm, which are 1.9-3.5x lower than that obtained with the FTB.

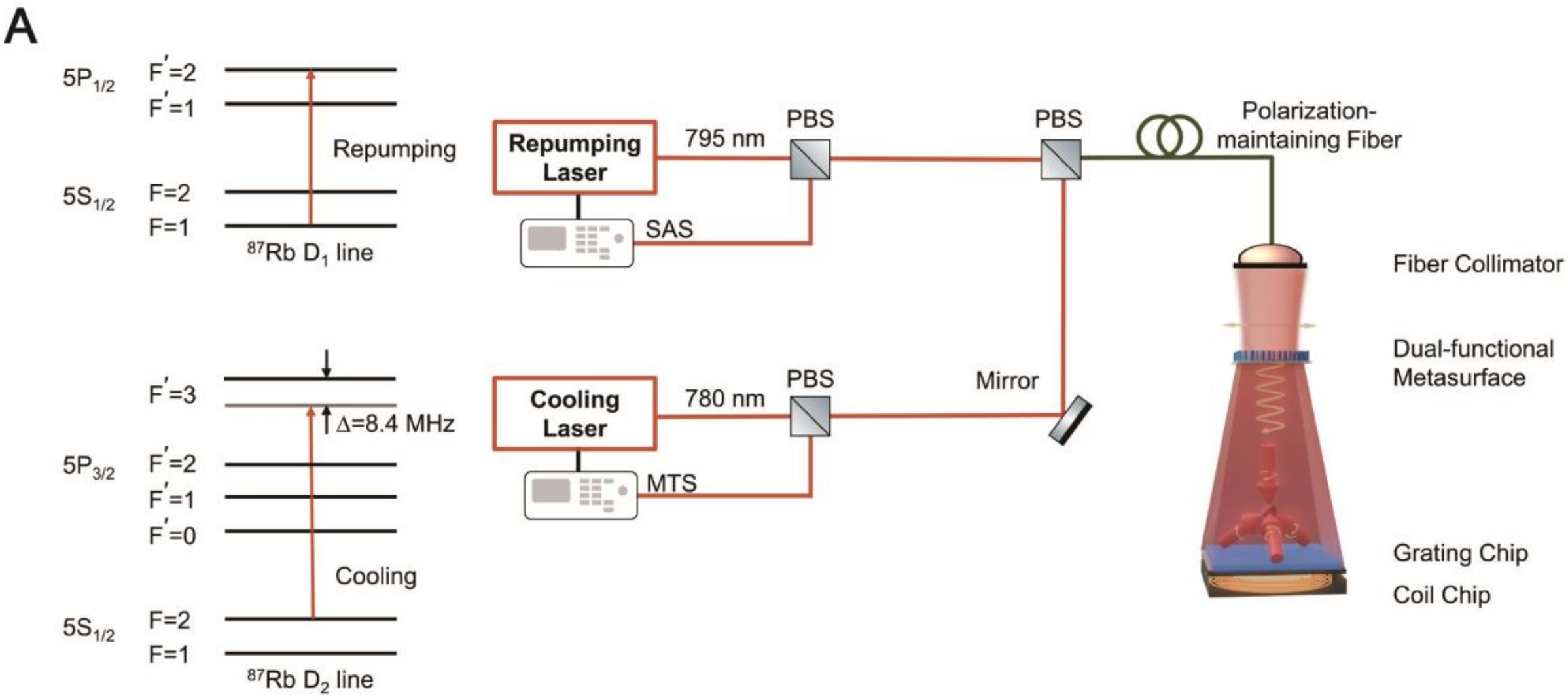

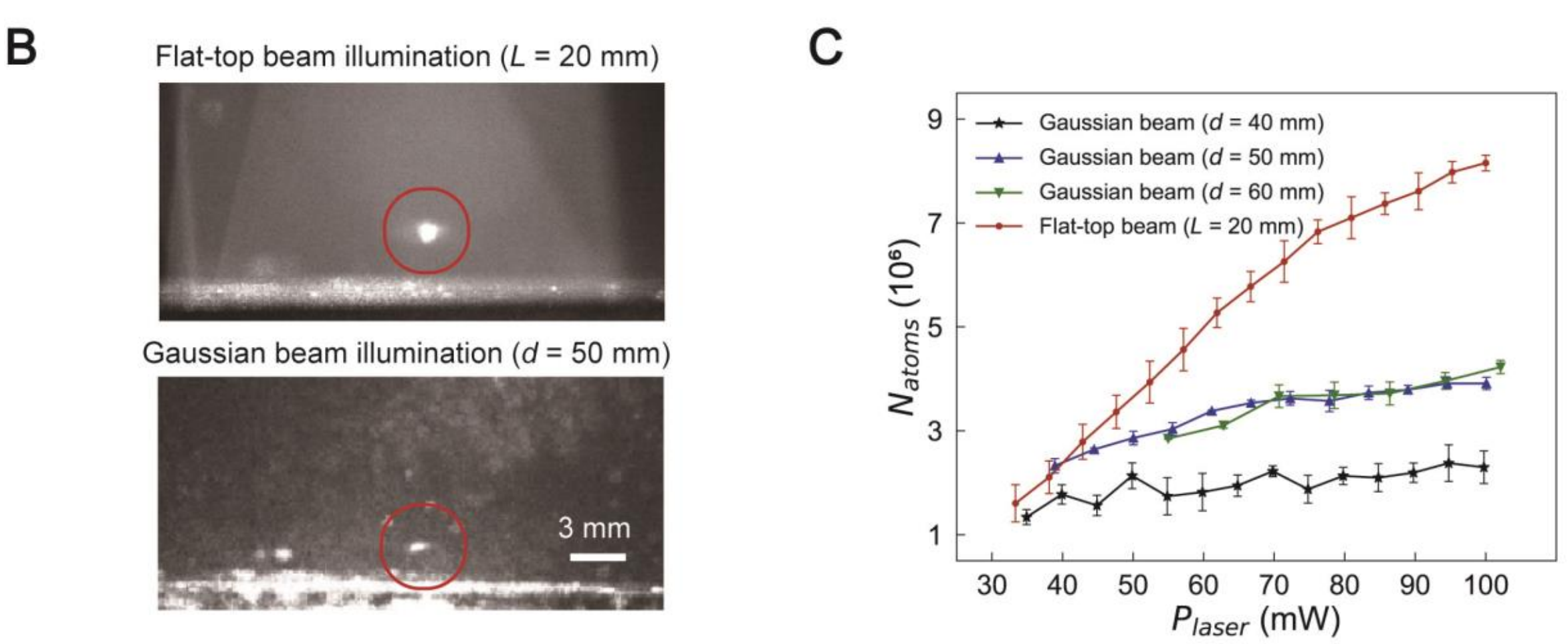

**Fig. 4. Experimental characterization of the low-SWaP GMOT system for trapping $^{87}$Rb atoms.** **(A)** Schematic of the energy levels of $^{87}$Rb atoms and the laser cooling setup. Left panel: $D_1$ and $D_2$ line of $^{87}$Rb, respectively. The 780 nm cooling laser is tuned to the $5S_{1/2}$ (F = 2) → $5P_{3/2}$ (F' = 3) cycling transition on the $D_2$ line, and the repumping laser is resonant with the $5S_{1/2}$ (F = 1) → $5P_{3/2}$ (F' = 2) transition. Right panel: Schematic illustration of the cooling setup. The cooling light is locked with the modulation transfer spectrum (MTS) technique, and the repumping light is stabilized with saturated absorption spectrum (SAS) of $^{87}$Rb. Both linearly-polarized laser beams are coupled into a polarization-maintaining fiber and subsequently out-coupled into free space by a fiber collimator. The dual-functional metasurface reshapes the linearly-polarized output Gaussian beam into a circularly-polarized square-shaped FTB which subsequently illuminates the planar grating chip. **(B)** Fluorescence images of trapped atoms under different illumination conditions. Upper panel: When an FTB (side length $L$ = 20 mm) is used, the trapped-atom cluster exhibits a circular shape. Lower panel: When an expanded Gaussian beam (diameter $d$ = 50 mm) is used, the trapped-atom cluster displays a flattened shape. **(C)** Measured trapped-atom number versus laser power for the cases of FTB and Gaussian beam illuminations. Error bars denote one standard deviations of the measured data.

We have demonstrated that the planar-component-based GMOT system not only delivers superior atom trapping performance—achieving 1.9-3.5x higher atom counts at the same laser power compared to a conventional GMOT with expanded Gaussian beam—but also reduces coil power consumption by more than two orders of magnitude. In the following, we show that this planar strategy further enables substantial reductions in both the weight and size of the system.

In the optical subsystem of the conventional GMOT setup in this study, a quarter waveplate is employed to convert the beam's polarization state to circular, and a pair of lenses is employed to expand the beam. Together, these optical components weigh 13.29 g and occupy a volume of 6.55 $cm^3$ (details elaborated in Section IX, Supplementary Materials). In contrast, the dual-functional metasurface—fabricated on a square substrate with a side length of 29.82 mm and thickness of 0.52 mm—weighs 1.01 g and occupies a volume of 0.46 $cm^3$, corresponding to a 13-fold reduction in weight and a 14-fold reduction in volume. Furthermore, when considering only the effective patterned area of the metasurface (1.50 mm × 1.50 mm square), the reduction increases to 4500-fold in weight and 1600-fold in volume.

The magnetic subsystem also benefits substantially from the planar design. The coil chip weighs 8.95 g — less than 1% of the weight of the wires in conventional anti-Helmholtz coils (1174.94 g). It is worth noting that the additional weight of the coil holder has been excluded from this comparison (details in Section IX, Supplementary Materials). While conventional anti-Helmholtz coils occupy 4084.07 $cm^3$, the planar coil chip has a total thickness of 2 mm and a footprint of 39 × 39 mm, corresponding to a volume of 3.04 $cm^3$ and representing a reduction of three orders of magnitude. A complete SWaP comparison between conventional and planar-component-based GMOT systems is provided in Table 1.

**Table 1. SWaP analysis of conventional and planar-component-based GMOT systems.**

| Systems | Optical subsystem | | | | Magnetic subsystem | | | |
|---|---|---|---|---|---|---|---|---|
| | Component | Diameter or Side length (mm) | Thickness (mm) | Weight (g) | Component | Volume ($cm^3$) | Weight (g) | Operating power (W) |
| Conventional GMOT | QWP | 25.33 | 3.20 | 4.003 | Anti-Helmholtz coil pairs | 4084.07 | 1174.94 | 67.00 |
| | Lens 1 | 25.03 | 7.25 | 6.069 | | | | |
| | Lens 2 | 25.38 | 2.71 | 3.213 | | | | |
| Planar-component-based GMOT | Metasurface (Considering the square substrate) | 29.82 | 0.52 | 1.009 | Planar coil chip | 3.12 | 8.95 | 0.56 |
| | Metasurface (Considering only the patterned area) | 1.50 | 0.52 | 0.003 | | | | |

## Conclusion

In summary, we have developed a fully planar, chip-based cold atom platform with significantly reduced system size, weight, and power consumption, by replacing conventional bulky optical and magnetic components with a dielectric dual-functional metasurface and a planar coil chip. The metasurface efficiently transforms a linearly-polarized Gaussian beam into a circularly-polarized flat-top beam, achieving both high intensity uniformity and degree of circular polarization. Meanwhile, the planar coil chip reliably generates the required quadrupole magnetic field for effective atom trapping. Experimental results demonstrate that the platform can trap up to $8.15 \times 10^6$ $^{87}$Rb atoms, achieving superior trapping efficiency over setups using expanded Gaussian beams. The seamless integration of both planar metasurface optics and planar magnetic component represents a major step toward the miniaturization of cold atom systems. Moreover, as atomic clocks and quantum sensors often require additional components to enhance signal-to-noise ratio (*69-72*)—adding system complexity and power demand—this on-chip integration strategy offers a compelling balance of performance, simplicity, and scalability. The proposed platform provides a foundation for the development of miniaturized and energy-efficient quantum technologies, enabling new possibilities for quantum sensing and precision measurement in field-deployable and resource-constrained environments such as compact atomic clocks, portable quantum sensors, and spacecraft-based quantum instruments.

## Materials and Methods

**Metasurface fabrication:** The metasurface fabrication process begins with the deposition of 1000-nm-thick amorphous silicon (a-Si) layer onto fused silica substrate using plasma-enhanced chemical vapor deposition (PECVD). The film's refractive index and thickness value are characterized by reflection-mode spectroscopic ellipsometry using the interference enhancement method (*73*), at three different angles of incidence (55°, 65°, and 75°) with respect to the normal to the plane of the a-Si layer. Electron beam (e-beam) lithography is employed to expose the metasurface pattern on a layer of positive-tone e-beam resist. After resist development, a 50-nm-thick chromium (Cr) layer is deposited onto the exposed resist layer, and a lift-off process is used to transfer the metasurface pattern onto the Cr layer, forming a hard mask for the subsequent Si etching. The metasurface pattern is then transferred from the Cr layer to the Si layer using an inductively coupled plasma reactive ion etching (ICP-RIE) process. Finally, the array of Si nanopillars is obtained by removing the Cr hard mask and cleaning the sample using solvent.